\input harvmac
\input epsf

\newcount\figno
\figno=0
\def\fig#1#2#3{
\par\begingroup\parindent=0pt\leftskip=1cm\rightskip=1cm\parindent=0pt
\baselineskip=11pt
\global\advance\figno by 1
\midinsert
\epsfxsize=#3
\centerline{\epsfbox{#2}}
\vskip 12pt
{\bf Fig. \the\figno:} #1\par
\endinsert\endgroup\par
}
\def\figlabel#1{\xdef#1{\the\figno}}
\def\encadremath#1{\vbox{\hrule\hbox{\vrule\kern8pt\vbox{\kern8pt
\hbox{$\displaystyle #1$}\kern8pt}
\kern8pt\vrule}\hrule}}

\overfullrule=0pt

\noblackbox
\parskip=1.5mm
%\def\semi{;~}

%%%%%%%%%%%%%%%%%%%%%%%%%%%%%%%%%%%%%%%%%%%%%%%%%%%%%%%%%%%%%%%%%%%
%%%  modify title page
%%%%%%%%%%%%%%%%%%%%%%%%%%%%%%%%%%%%%%%%%%%%%%%%%%%%%%%%%%%%%%%%%%%
\def\Title#1#2{\rightline{#1}\ifx\answ\bigans\nopagenumbers\pageno0
\vskip0.5in
\else\pageno1\vskip.5in\fi \centerline{\titlefont #2}\vskip .3in}

%\def\listrefs{\footatend\bigskip\bigskip\immediate\closeout\rfile
%\writestoppt \baselineskip =13pt\centerline{{\secfont References}}
%\bigskip{\frenchspacing\parindent =20pt \escapechar +'
%\input\jobname.refs \vfill\eject}\nonfrenchspacing} 
%%%%%%%%%%%%%%%%%%%%%%%%%%%%%%%%%%%%%%%%%%%%%%%%%%%%%%%%%%%%%%%%%%%%%%%%%%%%

\noblackbox
\parskip=1.5mm
%\def\semi{;~}

%%%%%%%%%%%%%%%%%%%%%%%%%%%%%%%%%%%%%%%%%%%%%%%%%%%%%%%%%%%%%%%%%%%%%
  
\def\npb#1#2#3{{\it Nucl. Phys.} {\bf B#1} (#2) #3 }
\def\plb#1#2#3{{\it Phys. Lett.} {\bf B#1} (#2) #3 }
\def\prd#1#2#3{{\it Phys. Rev. } {\bf D#1} (#2) #3 }
\def\prl#1#2#3{{\it Phys. Rev. Lett.} {\bf #1} (#2) #3 }

\def\ijmpa#1#2#3{{\it Int. J. Mod. Phys.} {\bf A#1} (#2) #3 }
\def\jmp#1#2#3{{\it J. Math. Phys.} {\bf #1} (#2) #3 }
\def\cmp#1#2#3{{\it Commun. Math. Phys.} {\bf #1} (#2) #3 }

\def\bb#1{{\tt hep-th/#1}}

\def\heph#1{{\tt hep-ph/#1}}
\def\mathph#1{{\tt math-ph/#1}}

\def\npps#1#2#3{{\it Nucl. Phys. Proc. Suppl. } {\bf #1} (#2) #3 }

\def\jhep#1#2#3{{\it JHEP} {\bf #1} (#2) #3 }
\def\cm#1{{\tt cond-mat/#1}}

%%%%%%%%%%%%%%%%%%%%%%%%%%%%%%%%%%%%%%%%%%%%%%%%%%%%%%%%%%%%%%%%%%%%%
%%%%%%%%%%%%%%%%%%%%    some definitions    %%%%%%%%%%%%%%%%%%%%%%%%%
%%%%%%%%%%%%%%%%%%%%%%%%%%%%%%%%%%%%%%%%%%%%%%%%%%%%%%%%%%%%%%%%%%%%%

           \def\CO{{\cal O}} 
  \def\CF{{\cal F}} 
\def\CL{{\cal L}}   
  \def\CD{{\cal D}} 
\def\CM{{\cal M}}  
\def\CN{{\cal N}}

\def\stackrel#1#2{\mathrel{\mathop{#2}\limits^{#1}}}

%%%%%%%%%%%%%%%%%%%%%%%%%%%%%%%%%%%%%%%%%%%%%%%%%%%%%%%%%%%%%%%%%%%%%

\def\dj{\hbox{d\kern-0.347em \vrule width 0.3em height 1.252ex depth
-1.21ex \kern 0.051em}}

\def\half{{1\over 2}\,}

%%%%%%%%%%%%%%%%%%%%%%%%%%%%%%%%%%%%%%%%%%%%%%%%%%%%%%%%%%%%%%%%%%%%%%
%%%%%%%%%%%%%%%%%%%%%%%        references         %%%%%%%%%%%%%%%%%%%%%%
%%%%%%%%%%%%%%%%%%%%%%%%%%%%%%%%%%%%%%%%%%%%%%%%%%%%%%%%%%%%%%%%%%%%%%%dsky

\lref\rzac{D.B. Fairlie, P. Fletcher and C.K. Zachos, \jmp{31}{1990}{1088.}}
\lref\mac{A.J. Macfarlane, A. Sudbery and P.H. Weisz, \cmp{11}{1968}{77.}}
\lref\kphd{T. Krajewski, {\it G\'eom\'etrie non commutative et interactions
fondamaentales}, Ph.D. Thesis. (\mathph{9903047})}
\lref\rj{M.M. Seikh-Jabbari, \jhep{06}{1999}{015.} (\bb{9903107})}
\lref\rkw{T. Krajewski and R. Wulkenhaar, {\it Perturbative quantum gauge fields
on the noncommutative torus}, Preprint CPT-99/P.3794. (\bb{9903187})}
\lref\rthooft{G. 't Hooft, \npb{153}{1979}{141;} \cmp{81}{1981}{267.}}
\lref\kap{J.I. Kapusta, \npb{148}{1979}{461.}}
\lref\bs{D. Bigatti and L. Susskind, {\it Magnetic fields, branes and 
nocommutative geometry}, Preprint SU-ITP-99-39. (\bb{9908056})}
\lref\rncym{T. Filk, \plb{376}{1996}{53.}}
\lref\rmagf{P. van Baal, \cmp{85}{1982}{529\semi}
J. Troost, {\it Contant field strenghts on $T^{2n}$}, Preprint VUB-TENA-99-04.
(\bb{9909187})}
\lref\br{J.L.F. Barb\'on and E. Rabinovici, {\it On 1/N corrections 
to the entropy of noncommutative gauge theories}, Preprint RI-15-99. 
(\bb{9910019})}
\lref\sw{N. Seiberg and E. Witten, \jhep{09}{1999}{032.} (\bb{9908142})}
\lref\niels{T. Harmark and N. Obers, {\it Phase structure of field theories
and spinning brane bound states}, Preprint NBI-HE-99-47. (\bb{9911169})}
\lref\rmr{J.M. Maldacena and J.G. Russo, \jhep{09}{1999}{25}. (\bb{9908134})}
\lref\rhio{A. Hashimoto and N. Itzhaki, \plb{465}{1999}{142.} (\bb{9907166})}
\lref\rcds{A. Connes, M. Douglas and A. Schwarz, \jhep{02}{1998}{003}. (\bb{9711162})}
\lref\wsft{E. Witten, \npb{268}{1986}{253.}}
\lref\rps{B. Pioline and A. Schwarz, \jhep{08}{1999}{021.} (\bb{9908019})}
\lref\rdh{M.R. Douglas and C.M. Hull, \jhep{02}{1998}{008.} (\bb{9711165})}
\lref\qhe{J. Bellisard, A. van Elst and H. Schulz-Baldes, {\it The non-commutative
geometry of the quantum Hall effect}, \cm{9411052}.}
\lref\ac{A. Connes and J. Lott, \npps{18}{1990}{29.}}
\lref\vm{M.A. V\'azquez-Mozo, \prd{60}{1999}{106010.} (\bb{9905030})}
\lref\ft{A. Fotopoulos and T.R. Taylor, \prd{59}{1999}{061701.} (\bb{9811224})}
\lref\sjr{C. Kim and S.-J. Rey, {\it Thermodynamics of large-N super Yang-Mills and the
AdS/CFT correspondence}, Preprint SNUST-99-005 (\bb{9905205}).}
\lref\an{A. Nieto and M.H.G. Tytgat, {\it Effective field theory approach to N=4 
supersymmetric Yang-Mills at finite temperature}, Preprint CERN-TH-99-153 (\bb{9906147}).}
\lref\raz{P. Arnold and C. Zhai, \prd{50}{1994}{7603} (\heph{9408276});
\prd{51}{1995}{1906} (\heph{9410360})\semi
C. Zhai and B. Kastening, \prd{52}{1995}{7232} (\heph{9507380}).}
\lref\rcb{A. Connes, {\it Noncommutative Geometry}, Academic Press 1994.}
\lref\rcar{C.P. Mart\'{\i}n and D. S\'anchez-Ruiz, \prl{83}{1999}{476.} (\bb{9903077})}
\lref\rtoi{T. Toimela, \plb{124}{1983}{407.}}
\lref\rbn{E. Braaten and A. Nieto, \prd{53}{1996}{3421.} (\bb{9510408})}
\lref\ob{T. Harmark and N.A. Obers, {\it Phase structure of noncommutative
field theories and spinning brane bound states}, Preprint NBI-HE-99-47 (\bb{9911169}).}
\lref\arlh{A. Connes, {\it Non-commutative geometry and physics}, in: "Gravitation and 
Quantizations", Proceedings of the 1992 Les Houches Summer School. Eds. B. Julia and
J. Zinn-Justin. Elsevier 1995.}
\lref\rpp{J.C. V\'arilly and J.M. Gracia-Bond\'{\i}a, \ijmpa{14}{1999}{1305.} (\bb{9804001})} 
\lref\rhid{A. Hashimoto and N. Itzhaki, {\it On the hierarchy between non-commutative and
ordinary supersymmetric Yang-Mills}, Preprint NSF-ITP-99-133, (\bb{9911057}).}
\lref\aoj{M. Alishahiha, Y. Oz and M.M. Seikh-Jabbari, {\it 
Supergravity and large N noncommutative field theories}, \jhep{11}{1999}{007.}
(\bb{9909215})}
\lref\rhaya{M. Hayakawa, {\it Perturbative analysis on infrared aspects of noncommutative
QED on ${\bf R}^{4}$}, (\bb{9912094}).}
\lref\rp{S. Minwalla, M. Van Raamsdonk and N. Seiberg, {\it Noncommutative perturbative
dynamics}, Preprint PUPT-1905, (\bb{9912072}).}
\lref\aref{I.Ya. Aref'eva, D.M. Belov and A.S. Koshelev, {\it 
Two-loop diagrams in noncommutative $\phi^4_{4}$ theory}, Preprint
SMI-15-99 (\bb{9912075}).}
\lref\rgc{H. Garc\'{\i}a-Compean, \npb{541}{1999}{651.} (\bb{9804188})}
\lref\rcai{R.-G. Cai and N. Ohta, {\it On the thermodynamics of large-N noncommutative super
Yang-Mills theories}, Preprint OU-HET-329 (\bb{9910092}).}
\lref\rhv{C. Hofman and E. Verlinde, \jhep{12}{1998}{010.} (\bb{9810116})}
\lref\rcor{R. Parwani and C. Corian\`o, \npb{434}{1995}{56} (\heph{9409069});
\prl{73}{1994}{2398} (\heph{9405343}).}
\lref\rrome{M. Bianchi, G. Pradisi and A. Sagnotti, \npb{376}{1991}{365\semi}
A. Sen and S. Sethi, \npb{499}{1997}{45} (\bb{9703157})\semi
M. Bianchi, \npb{528}{1998}{73} (\bb{9711201})\semi
E. Witten, \jhep{02}{1998}{006} (\bb{9712028}).}

%%%%%%%%%TEXT%%%%%%%%%%%%%%%%%%%%%%%%%%%%%%%%%%%%%%%%%%%%%%%%%%%%%%%%%
%%%%%%%%%%%%%%%%%%%%%%%%%%%%%%%%%%%%%%%%%%%%%%%%%%%%%%%%%%%%%%%%%%%%%B
%%%%%%%%%%%%%%%%%%          title page       %%%%%%%%%%%%%%%%%%%%%%%%%
%%%%%%%%%%%%%%%%%%%%%%%%%%%%%%%%%%%%%%%%%%%%%%%%%%%%%%%%%%%%%%%%%%%%%%

\line{\hfill SPIN-1999/31}
\line{\hfill DFTT-99-66}
\line{\hfill ITFA-99-41}  

\line{\hfill {\tt hep-th/9912140}}
%\vskip 0.1cm

\Title{\vbox{\baselineskip 12pt\hbox{}
 }}
{\vbox{{\centerline{Thermal effects in perturbative }
{\centerline{     }}
{\centerline{noncommutative gauge theories}}
}}}

%\vskip 0.1cm

\centerline{$\quad$ {G. Arcioni$^{\rm \,a,c,\,}$\foot{E-mail: Arcioni@to.infn.it, 
G.Arcioni@phys.uu.nl}
and
M.A. V\'azquez-Mozo$^{\rm \,b,c,\,}$\foot{E-mail: vazquez@wins.uva.nl, M.Vazquez-Mozo@phys.uu.nl}
 }}

\vskip 1cm

\centerline{{\sl $^{\rm a\,}$ Dipartimento di Fisica Teorica, Universit\`a di Torino,}}
\centerline{{\sl Via P. Giuria 1, I-10125 Torino, Italy}}

%\centerline{{and}}

\centerline{{\sl $^{\rm b\,}$ Instituut voor Theoretische Fysica, Universiteit van Amsterdam,}} 
\centerline{{\sl Valckenierstraat 65, 1018 XE Amsterdam, The Netherlands}}

%\centerline{{and}}

\centerline{{\sl $^{\rm c\,}$ Spinoza Instituut, Universiteit Utrecht,}}
\centerline{{\sl Leuvenlaan 4, 3584 CE Utrecht, The Netherlands}}

 \vskip 1.2cm

\noindent  
The thermodynamics of gauge theories on the noncommutative plane is studied 
in perturbation theory. For $U(1)$ noncommutative Yang-Mills we compute the
first quantum correction to the ideal gas free energy density and study their behavior
in the low and high temperature regimes. Since the noncommutativity scale 
effectively cutoff interactions at large distances, the theory is regular 
in the infrared. In the case of $U(N)$ noncommutative Yang-Mills we evaluate
the two-loop free energy density and find that it depends on the noncommutativity 
parameter through the contribution of non-planar diagrams.

%%%%%%%%%%%%%%%%%%%%%%%%%%%%%%%%%%%%%%%%%%%%%%%%%%%%%%%%%%%%%%%%%%%%%%

\Date{12/99}
%\draft

%%%%%%%%%%%%%%%%%%%%%%%%%%%%%%%%%%%%%%%%%%%%%%%%%%%%%%%%%%%%%%%%%%%%%%%%%%
%%%%%%%%%%%%                text begins                        %%%%%%%%%%%
%%%%%%%%%%%%%%%%%%%%%%%%%%%%%%%%%%%%%%%%%%%%%%%%%%%%%%%%%%%%%%%%%%%%%%%%%%

\newsec{Introduction}

Noncommutative geometry \refs\rcb\ has been a recurrent issue in physics in the last decades. 
After several attempts to incorporate the mathematical formalism in string field theory 
\wsft\ and
even the standard model \refs\ac\refs\arlh\ and the quantum Hall effect \refs\qhe, it has 
recently re-emerged in the context of string/M-theory in the presence of constant 
background fields \refs\rcds\refs\rdh\refs\sw. Since the low-energy limit of these 
configurations is described in terms of a supersymmetric gauge theory living in a noncommutative
space, the study of these type of nonlocal field theories has received renewed attention
lately. Now, however, because of their stringy connections, new tools are available to 
study the physics of noncommutative field theories. As a matter of example,  
the extension of the AdS/CFT correspondence to backgrounds with constant vacuum values of the
(Neveu-Schwarz)$^2$ tensor field \refs\rhio\refs\rmr\ makes it feasible the study 
of noncommutative field theories also in the strong coupling regime 
\refs\aoj\refs\br\refs\rcai\refs\rhid\refs\ob. On the perturbative side, 
several aspects of noncommutative gauge theories have been recently 
addressed in \refs\rp\refs\aref\refs\rhaya.

The physical idea behind the application of noncommutative geometry is that of the
quantization of space-time itself by introducing noncommuting space-time
coordinates
\eqn\crel{
[x^{\mu},x^{\nu}]=2i\theta^{\mu\nu}, \hskip 1cm \mu,\nu=0,\ldots,d-1.
}
Roughly speaking, in ordinary quantum theory the canonical commutation relations lead
to a quantization of the phase space that results in a smeared symplectic geometry
at short distances due to the uncertainty principle. Following a similar line of
reasoning, one is lead to think that the commutation relations \crel\ will smear the
space-time picture at distances shorter than $\sqrt{\theta}$, imposing thus a natural
cutoff for the description of Nature in terms of a local quantum field theory.

In the formalism on noncommutative geometry, the geometrical features of the noncommutative
manifold are reconstructed by considering the deformation of the $C^{*}$-algebra of
continuous complex functions defined on it and vanishing at infinity, using the Weyl 
product
\eqn\moyal{
(f\star g)(x)=f(x)e^{i\stackrel{\leftarrow}{\partial_{i}}\theta^{ij}
\stackrel{\rightarrow}{\partial}_{j}}g(x).
}
Consequently, quantum field theories on noncommutative spaces can be formulated
by writing the ordinary action and replacing the commutative product with the
$\star$-product defined by \moyal; because of its non-polynomial character, 
the resulting field theories will be non-local. It is actually this non-locality 
what is argued to smear physics at short distances.

The study of systems at finite temperature usually provides good insights into their
physical behavior. In this note we will be concerned with the thermodynamics of 
Yang-Mills theories 
on noncommutative spaces (NCYM) of the type $\CM_{\theta}\times {\bf R}_{t}$ where 
$\CM_{\theta}$ is some $(d-1)$-dimensional noncommutative space, typically 
${\bf R}_{\theta}^{d-1}$, characterized by a deformation matrix $\theta^{ij}$ 
($i,j=1,\ldots,d-1$, $d>2$).
In this setup, we can compute the thermodynamical potentials using the imaginary
time formalism by compactifying the euclidean time to length $\beta=T^{-1}$. 
The corresponding Feynman rules are thus obtained from the Euclidean Feynman rules
of zero temperature noncommutative gauge theories by quantizing the time components 
of the momenta in units of $2\pi T$.
One technical payoff of restricting noncommutativity to the spatial sections is that the
resulting non-polynomial functions of the momenta in the Feynman integrals do not involve the 
discrete Euclidean momentum. Thus, the Matsubara sums that appear in the computation of 
the free energy are of the same kind that one encounters in ordinary commutative quantum
field theories. 

In the following we will focus our attention on gauge theories on ${\bf R}_{\theta}^{2}
\times {\bf R}_{t}$ and ${\bf R}_{\theta}^{3}\times {\bf R}_{t}$. Actually, in the four-dimensional
case we can always find a rigid orthogonal coordinate transformation $\tilde{x}=Ax$ that 
takes a generic antisymmetric matrix $\theta_{ij}$ to its block off-diagonal form
$$
A\left(\matrix{0 & \theta_{12} & \theta_{13} \cr
-\theta_{12} & 0 & \theta_{23} \cr
-\theta_{13} & -\theta_{23} & 0}\right) A^{T} =
\left(\matrix{0 & \theta & 0\cr -\theta & 0 & 0 \cr 
0 & 0 & 0 }\right)
$$
with $\theta^{2}=\theta_{12}^{2}+\theta_{13}^{2}+\theta_{23}^{2}$. Thus, when
expressed in the appropriate system of coordinates, we see that ${\bf R}_{\theta}^{3}
\times{\bf R}_{t}$ is actually equivalent to ${\bf R}_{\theta}^{2}\times {\bf R}\times
{\bf R}_{t}$.

The present paper is organized as follows: in Sec. 2 we study the loop corrections to the
thermodynamics of $U(1)$ pure NCYM theories at finite temperature. Sec. 3 will be devoted to
the study of the non-abelian $U(N)$ case (and its supersymmetric extensions), where we will 
find that all the dependence of the two-loop free energy density on $\theta$ comes from the contribution 
of the $U(1)$ part of $U(N)$ to non-planar diagrams. Finally, in Sec. 4 we will 
summarize our conclusions.

\newsec{Thermodynamics of $U(1)$ NCYM}

Noncommutative pure $U(1)$ Yang-Mills theory is specially interesting, since in this case
interaction appears solely as the result of noncommutativity. The action of a $U(1)$ gauge
field on ${\bf R}^{d-1}_{\theta}\times {\bf R}_{t}$ can be written as
\eqn\ncuo{
S_{U(1)} = -{1\over 4} \int d^{d}x\,F_{\mu\nu}\star F^{\mu\nu}
}
where the star product is defined by \moyal\ and the field strength is given 
in terms of the Moyal bracket $\{f,g\}_{\rm MB}=f\star g-g \star f$ as
$$
F_{\mu\nu}=\partial_{\mu}A_{\nu}-\partial_{\nu}A_{\mu}+ig\{A_{\mu},A_{\nu}\}_{\rm MB}.
$$

The Feynman rules for this theory are easily written in momentum space, as shown in
references \refs\rj\refs\rkw. The resulting diagrammatic expansion is qualitatively similar 
to ordinary non-abelian Yang-Mills theories, except for the momentum dependence of 
the vertices through the function $\sin (\theta_{ij}p^{i}q^{j})$. As we will see later, 
this extra dependence on the momenta with respect to the ordinary non-abelian Yang-Mills 
theory has important consequences on the infrared behavior of the noncommutative theory.

At one-loop level, the free energy density is determined by the quadratic
part of the action and as a consequence it is independent of the noncommutativity of the base
space. Thus, the result is identical of that of pure QED$_{d}$, namely
\eqn\ol{
{\cal F}(T)_{\rm 1-loop}=-(d-2){\Gamma(d/2)\over \pi^{d\over 2}}\zeta(d)T^{d}.
}

Corrections to the ideal gas contributions can be computed in perturbation theory using the 
Feynman rules given in ref. \refs\rkw\ (see also \refs\kphd). The first term correcting
eq. \ol\ comes from two-loop diagrams. In our case, the final result can be cast in the form
\eqn\tl{
\CF(T)_{\rm 2-loop} = g^{2}(d-2)^{2}\int {d^{d-1}p\over (2\pi)^{d-1}}\int {d^{d-1}q\over (2\pi)^{d-1}}
{n_{b}(p)n_{b}(q)\over \omega_{p}\omega_{q}}\sin^{2}\theta(p,q)
}
where $\omega_{p}=p$, $\theta(p,q)=\theta_{ij}p^{i}q^{j}$ and 
$$
n_{b}(p)={1\over e^{p/T}-1}
$$
is the Bose-Einstein distribution function. Ultraviolet divergences in the zero temperature
part of the diagrams are taken care of by inserting the corresponding one-loop couterterms at
$T=0$ \refs\rkw\refs\rcar.

A first thing to be noticed is that, in spite of the similarities between the diagrammatic expansion
of the free energy of noncommutative $U(1)$ gauge theory and that of pure YM$_{d}$, here the 
infrared behavior is much softer due to the presence of the factor $\sin^2\theta(p,q)$.
which when $p\rightarrow 0$ will vanish as $\CO(p^2)$. Thus, the two-loop correction \tl\ is
well defined for all $d\geq 3$.

Let us first analyze the three dimensional case ($d=3$). Here, the antisymmetric matrix $\theta_{ij}$
can be written as $\theta_{ij}=\theta \epsilon_{ij}$ and the integration over angular variables in
\tl\ can be easily performed with the result
\eqn\td{
\CF(T)_{\rm 2-loop}={g^2T^2 \over 8\pi^2}
\int_{0}^{\infty}du\int_{0}^{\infty}dv\,{1-J_{0}(2\theta T^2 uv)\over 
(e^{u}-1)(e^{v}-1)}.
}
We notice that this integral is both infrared ($u,v \rightarrow 0$) and ultraviolet
($u,v\rightarrow \infty$) convergent. Let us study first the case when the temperature
is much smaller than the energy scale $1\over \sqrt{\theta}$ associated with noncommutativity
effects. If $T\sqrt{\theta}\ll 1$, we can expand the 
Bessel function in power series and integrate term by term. The result is an asymptotic 
series valid for small $T\sqrt{\theta}$ whose first term is 
$$
\CF(T)_{\rm 2-loop}\sim {\zeta(3)^2\over 2\pi^2}g^2\theta^2T^6 +\ldots
$$
The asymptotic character of the series is easily understood by realizing that by expanding 
the Bessel function in power series and truncating the series we fall short in reproducing 
the integrand in the ultraviolet region, but this is precisely the region of the integral that is
effectively cutoff at low temperatures.

The evaluation of \td\ in the opposite region $T\gg 1/\sqrt{\theta}$ is more complicated due to
the peculiar infrared structure of the theory at hand. We know that, asymptotically, the 
Bessel function oscillates very fast for large values of the argument. If we introduce 
a cutoff $\Lambda_{\theta}$ (that in principle will depend on the value of $T\sqrt{\theta}$)
to isolate the infrared sector of the integral, we have
that
$$
\int_{\Lambda_{\theta}}^{\infty}du\int_{\Lambda_{\theta}}^{\infty}dv\,{1-J_{0}(2\theta T^2 uv)
\over (e^{u}-1)(e^{v}-1)}\sim \int_{\Lambda_{\theta}}^{\infty}{du\over e^{u}-1} 
\int_{\Lambda_{\theta}}^{\infty}
{dv\over e^{v}-1}
$$
since for large values of the argument the rapidly oscillating Bessel function will 
be averaged to zero. Thus, when $T\sqrt{\theta}\gg 1$ we can write 
\eqn\dec{
\CF(T)_{\rm 2-loop} \sim {1\over 3}\CF(T,\Lambda_{\theta})^{SU(2)}_{\rm 2-loop}+ 
f(T,\Lambda_{\theta})_{\rm IR}
}
where $\CF(T,\Lambda_{\theta})^{SU(2)}_{\rm 2-loop}$ is the two-loop free energy density
of ordinary
pure YM$_{3}$ with gauge group $SU(2)$ and infrared momentum cutoff $\Lambda_{\theta} T$, and 
$f(T,\Lambda_{\theta})_{\rm IR}$ is some contribution containing the information from the infrared sector of the theory.

We can interpret the decomposition \dec\ in the sense that, in the large-$T\sqrt{\theta}$ limit,
the ultraviolet sector of the $U(1)$ noncommutative gauge theory is well described in terms
of an ordinary Yang-Mills theory with $C_{2}(G)=2$ \refs\rcar\refs\rj\refs\rkw. 
The numerical prefactor ${1\over 3}$ just
reflects the fact that the $U(1)$ noncommutative gauge theory only has one propagating ``photon''
while the $SU(2)$ Yang-Mills theory has three propagating gluons. Thus, the contribution 
to the free energy from the ultraviolet part of the theory is given by the free energy {\it
per vector boson} of an $SU(2)$ commutative gauge theory. On the other hand, we see that the theory
in the infrared is radically different from ordinary Yang-Mills which is infrared divergent
at two loops in three dimensions. This feature is very much reminiscent of the Morita equivalence
\refs\rps\ between $U(1)$ Yang-Mills theory on the noncommutative torus and a $U(N)$ gauge theory on 
a commutative one in the presence of a magnetic flux, where the infrared sector of the theory
is regularized by the presence of the background twisted gauge field \refs\rthooft\refs\rmagf.

Let us focus our attention now on the four-dimensional case. From the discussion in the Introduction,
we know that we can choose coordinates $x,y,z$ such that noncommutativity is restricted to the 
$xy$-plane, $[x,y]=2i \theta$, $[x,z]=[y,z]=0$. Since the interacting character of the theory is
entirely due to the noncommutativity of the base space, we find that our theory will be free
whenever the momenta are orthogonal to the $xy$-plane. Thus, in order to study the integral
\td\ it is convenient to use cylindrical coordinates where the $z$-coordinate coincides with the 
``central'' direction. Again we can integrate over angular variables to find
\eqn\fdtl{
\eqalign{
\CF(T)_{\rm 2-loop} &= {g^2 T^4 \over 8\pi^4}\int_{-\infty}^{\infty}du_{z}\int_{-\infty}^{\infty}
dv_{z} \cr 
\times &
\int_{0}^{\infty}{u du\over \sqrt{u^2+u_{z}^2}}\int_{0}^{\infty}{v dv\over 
\sqrt{v^2+v_{z}^2}}{1-J_{0}(2T^2\theta\,uv)\over (e^{\sqrt{u^2+u^2_{z}}}-1)(e^{\sqrt{v^2+
v_{z}^2}}-1)}
}
}
As in three dimensions, when $T\sqrt{\theta}$ is small we can expand the Bessel function to 
get an asymptotic series in powers of $T\sqrt{\theta}$,
$$
\CF(T)_{\rm 2-loop} \sim 2g^2\left({\pi^2\over 45}\right)^2\theta^2T^8+\ldots
$$
When $T\sqrt{\theta}\gg 1$ it is difficult to estimate the value of the integral \fdtl. Since
the function resulting from integration of the angular variables is the same than in the 
three-dimensional case, we can argue along similar lines that, again, the free energy can be 
decomposed as in \dec\ into a $SU(2)$-like ordinary Yang-Mills piece and a contribution that 
takes care of the infrared sector of the theory. Although ordinary Yang-Mills in four-dimensions
is infrared finite at two loops and each term in \dec\ is finite in the limit 
$\Lambda_{\theta}\rightarrow 0$ (when $T\sqrt{\theta}\rightarrow \infty$), 
the infrared part $f(T,\Lambda_{\theta})_{\rm IR}$ gives always a non-trivial contribution, even 
in this limit.
This is due to the fact that in the infrared region the function multiplying $J_{0}(2T^2\theta uv)$
in the integrand is unbounded and rapidly varying close to $u,v=0$. Thus, when $T\sqrt{\theta}$ is very 
large the highly oscillatory Bessel function is not able to average it to zero since the function 
multiplying it is not approximately constant in a single period.

In principle, one could compute higher order corrections to the thermodynamic potential \fdtl. 
In ordinary YM$_{4}$ at finite temperature
the self-interaction of gluons introduce infrared divergences at
three loops that do not cancel order by order in perturbation theory and have to be taken 
care of by resumming the so-called ring diagrams. The result is a mild breakdown of the 
perturbative expansion that now is no longer a series in $g^2$ alone but also contains terms of 
order $g^3$ \refs\kap, $g^4 \log g^2$ \refs\rtoi\ and $g^5$ \refs\raz\refs\rbn (for similar results
in QED see \refs\rcor). The situation is worsened by extra divergences due to the self-interaction of the 
transverse gluons that invalidates perturbation theory at $\CO(g^6)$.

We can study what happens with $U(1)$ Yang-Mills theory on ${\bf R}_{\theta}^{2}\times 
{\bf R}\times {\bf R}_{t}$ for higher loops contributions to the free energy. A first problem 
to be solved would be whether the noncommutative $U(1)$ theory itself is renormalizable at $T=0$ beyond
one loop. Let us however assume that the ultraviolet divergences in the zero temperature sector
can be handled by some cutoff and concentrate our attention on the (ultraviolet finite) 
temperature dependent contributions. At three loops, infrared divergences are associated with the 
existence of a non-vanishing thermal mass at one loop. In the case at hand, however, if we compute the 
static limit of the one-loop self-energy of the gluon we find that it vanishes quadratically with the 
external spatial momentum,
$$
\Pi_{00}(q_{0}=0,\vec{q}\rightarrow 0)=2 \Pi_{\mu}^{\,\,\,\mu}(q_{0}=0,\vec{q}\rightarrow 0)\sim
{16\pi^2\over 45} g^2 \theta^2T^{4}q_{xy}^2 + \CO(q_{xy}^3)
$$
where $q_{xy}$ is the modulus of the projection of the external momentum $\vec{q}$ on the $xy$-plane. 
The first consequence of this fact, is that three-loop diagrams are free of infrared divergences and
therefore the next correction to the free energy is of order $g^{4}$. 

The soft behavior of $U(1)$ NCYM at low momenta (or large distances) follows from the fact
that in this limit the theory becomes free. In physical terms, this is because the theory should
reduce itself to its commutative version (a free $U(1)$ pure gauge theory) at length scales 
much bigger than the typical scale of noncommutative effects, i.e. $\sqrt{\theta}$. In this sense, 
this scale plays the role of an infrared cutoff for the noncommutativity-induced interactions, which
in the ultraviolet resemble those of a non-abelian gauge theory. Thus, the absence of infrared divergences
and non-analytic terms in $g^2$ in the lowest orders in perturbation theory seems to be a generic 
feature of the whole perturbative expansion which would be a series in integer powers of $g^2$.

Before closing this Section, let us make some remarks about the possible supersymmetric extensions
of $U(1)$ NCYM. In the commutative case, we can construct a trivial supersymmetric theory by adding
to the pure YM$_4$ theory the action of a free massless Majorana or Weyl spinor. This theory can be deformed into
an interacting supersymmetric theory by switching on the noncommutativity of the base 
space\foot{Extended supersymmetric theories can be obtained by considering a similar action in
dimension six or ten and performing dimensional reduction to four dimensions.}
\eqn\sncym{
S= \int d^{4}x \left[-{1\over 4} F_{\mu\nu}\star F^{\mu\nu}+i\bar{\psi}\star\gamma^{\mu}D_{\mu}
\psi\right],
}
where the covariant derivative is defined by $D_{\mu}\psi=\partial_{\mu}\psi-i(A_{\mu}\star\psi-
\psi\star A_{\mu})$. To compute the two-loop free energy density
now we have to add to the result for
pure NCYM the contribution coming from the fermion loop with the result
\eqn\ss{
\CF(T,\theta)_{\rm 2-loop}=4 g^2 \int {d^3 p\over (2\pi)^3}\int {d^3q\over 
(2\pi)^3}{1\over \omega_{p}\omega_{q}}[n_{b}(p)n_{b}(q)+2n_{b}(p)n_{f}(q) + 
n_{f}(p)n_{f}(q)]\sin^{2}\theta(p,q)
}
with $n_{f}(p)=1/(e^{p/T}+1)$ the Fermi-Dirac distribution function. The structure of expression
\ss\ is similar to the two-loop free energy of ordinary SYM \refs\vm.
Here again the theory at low momenta becomes trivial (a free gauge field plus an
``adjoint'' $U(1)$ fermion), rendering the theory infrared finite, while in the ultraviolet
it resembles $\CN=1$ SYM$_{4}$ with $C_{2}(G)=2$. The three-dimensional case can be 
worked out along similar lines, starting with the action \sncym\ in three dimensions (with
$\psi$ a Majorana spinor). As in the non-supersymmetric case, the result is
infrared finite.

\newsec{Non-abelian NCYM}

Let us consider now non-abelian four-dimensional noncommutative gauge theories on ${\bf R}_{\theta}^{2}
\times {\bf R}\times {\bf R}_{t}$. The non-abelian generalization of the action \ncuo\ is easily 
written as (group generators are normalized according to ${\rm Tr} \,T^{a}T^{b}=\half\delta^{ab}$)
\eqn\ncym{
S=-{1\over 2}\int d^{4}x \,{\rm Tr}\, [F_{\mu\nu}\star F^{\mu\nu}]
}
where $F_{\mu\nu}=\partial_{\mu}A_{\nu}-\partial_{\nu}A_{\mu}-ig(A_{\mu}\star A_{\nu}-A_{\nu}\star 
A_{\mu})$. Now since our fields are matrix-valued functions, the $\star$-product is defined by 
tensoring the Weyl product \moyal\ with the ordinary product of matrices. A first consequence 
of this is that in order for the gauge fields to form a closed algebra we should restrict the gauge
group to $U(N)$ \refs\sw, since we have to demand the group generators to form a closed algebra 
under ordinary matrix multiplication\foot{From the point of view of string theory, one can 
in principle allow for other groups such as $SO(N)$ and $USp(2N)$ by introducing orientifold
planes. Naively, the introduction of orientifolds will project out the (Neveu-Schwarz)$^2$
antisymmetric tensor field, and so it looks like one can only obtain commutative space-times. 
However, the orientation projection is compatible with having quantized background values of the 
(Neveu-Schwarz)$^2$ field \refs\rrome\ that will lead to a deformed product of functions on the
manifold. This deformed product, however, is not necessarily non-commutative. We
thank J. de Boer for a discussion on this point.}. From a purely 
quantum field theoretical point of view, one can try to construct noncommutative gauge theories using
Moyal brackets with gauge groups different from $U(N)$. At this level the obstruction to 
consider these other groups arises in the form of inconsistencies of the resulting non-local quantum 
field theory. For $SU(N)$, for example, the gauge variation of the
vector field $A_{\mu}$, $\delta A_{\mu}=\partial_{\mu}\lambda+i(\lambda\star A_{\mu}-A_{\mu}\star
\lambda)$, gets an extra piece proportional to the identity matrix
$$
\delta A_{\mu} = {i\over 2N}(\lambda^{a}\star A_{\mu}^{a}-A_{\mu}^{a}\star \lambda^{a}){\bf 1}+\ldots
$$
where the $\star$-product on the right hand side corresponds to that of ordinary functions. 
Thus, ordinary gauge transformations do not keep the vector field in the adjoint representation 
of $SU(N)$. Notice, however, that this extra piece scales as $\CO(1/N)$ and therefore disappears when 
$N\rightarrow \infty$ where the theory reduces itself to $U(N)$ in the large-$N$ limit, as well as
in the commutative limit when we recover ordinary $SU(N)$ YM.

In the ideal gas approximation, the free energy is easily computed
by adding the contribution of the different degrees of freedom. Since the noncommutativity parameter 
$\theta$ appears only in the interactions terms, the result is identical to the corresponding 
ordinary Yang-Mills theory
\eqn\tlna{
\CF(T)_{\rm 1-loop} = -{\rm dim}\,G \,{\pi^2\over 45}T^4
}
where ${\rm dim}\, G$ is the dimension of the gauge group, $N^2$ for $U(N)$.

In order to compute loop corrections to the free gas approximation, we need the Feynman rules for the 
non-abelian noncommutative Yang-Mills theory. Now, in contrast with the $U(1)$ case, there are 
interactions surviving the commutative $\theta\rightarrow 0$ limit, so the structure of the 
vertices will be more involved.  In order to make the computation more transparent, instead of 
using the ``trigonometric basis'' 
for the gauge group generators \refs\rzac, we will write Feynman rules using the structure 
constants $f^{abc}$ and Gell-Mann tensor $d^{abc}$ of the gauge group, defined in terms of the
generators $T^{a}$ by the identities
$$
\eqalign{
f^{abc}&= -2i\, {\rm Tr}\, [T^a,T^b]T^c \cr
d^{abc}&= 2 \,{\rm Tr}\, \{T^a,T^b\}T^c
}
$$
where by $\{\cdot,\cdot\}$ we represent the anticommutator of the two generators. 

The Feynman rules for the noncommutative $U(N)$ Yang-Mills can be easily obtained by 
writing the action \ncym\ in momentum space. It turns out that the only change with respect to
the Feynman rules of $U(N)$ ordinary Yang-Mills is the replacement on each vertex of the structure 
constants according to
\eqn\rep{
f^{a_{1}a_{2}a_{3}} \longrightarrow f^{a_{1}a_{2}a_{3}}\cos\theta(p_{1},p_{2})+ d^{a_{1}a_{2}a_{3}}
\sin\theta(p_{1},p_{2})
}
where $a_{i}$ is the color index associated with the particle with (incoming) momentum $p_{i}$. 
With this only change, the first quantum correction to \tlna\ can be  computed to give
\eqn\tlth{
\CF(T,\theta)_{\rm 2-loop} = g^2\int{d^{3}p\over (2\pi)^{3}}{n_{b}(p)\over \omega_{p}}\int{d^{3}q\over 
(2\pi)^{3}}
{n_{b}(q)\over \omega_{q}}\, \left[ f^{abc}f^{abc}\cos^2\theta(p,q)+d^{abc}d^{abc}\sin^2
\theta(p,q)\right].
}
However, for $U(N)$ ($N>1$) it turns out to be that\foot{For $N=1$ we recover the results of
the previous Section by setting $f^{aaa}=0$ and $d^{aaa}=2$.}
$$
f^{abc}f^{abc}= N(N^2-1), \hskip 1cm d^{abc}d^{abc}=N(N^{2}+1)
$$
and therefore the two-loop free energy density is  
$$
\eqalign{
\CF(T,\theta)_{\rm 2-loop} =& g^2N (N^2-1) \left[\int{d^{3}p\over (2\pi)^{3}}{n_{b}(p)\over 
\omega_{p}}\right]^2+2 g^2 N\int{d^{3}p\over (2\pi)^{3}}{n_{b}(p)\over \omega_{p}}\int{d^{3}q\over 
(2\pi)^{3}} {n_{b}(q)\over \omega_{q}}\sin^{2}\theta(p,q) \cr
=&{N^{2}-1\over 144}g^{2}NT^{4}+
2 g^2 N\int{d^{3}p\over (2\pi)^{3}}{n_{b}(p)\over \omega_{p}}\int{d^{3}q\over 
(2\pi)^{3}} {n_{b}(q)\over \omega_{q}}\sin^{2}\theta(p,q)
}
$$
The first ($\theta$-independent) part of $\CF(T,\theta)_{\rm 2-loop}$ is just the result for
the two-loop free energy of ordinary pure $SU(N)$ Yang-Mills theories in four dimensions 
\refs\kap. On the other hand, the $\theta$-dependence in the free energy density
is subleading in the large-$N$ limit and comes from the contribution of the $U(1)$ part 
of $U(N)$ to the two-loop non-planar diagrams. Incidentally, in the large-$T\sqrt{\theta}$ 
limit the contribution of the $U(1)$ part of $U(N)$ in the ultraviolet combines with
contribution of the $SU(N)$ part in the same region in such a way that the final result 
is of order $\CO(N^2)$ for finite $N$. This is a further evidence of the planar character of 
noncommutative Yang-Mills in the large-$\theta$ regime at fixed temperature (or at
large temperatures and fixed $\theta$) \refs\bs\refs\br\refs\ob\refs\rp.
 
In evaluating the free energy density at three loops, we are faced with
the customary infrared divergences coming from the ``commutative" $SU(N)$ part, since
the $U(1)$ part decouples at large distances ($p\ll 1/\sqrt{\theta}$) and will not induce 
infrared divergences. As usual, 
these infrared singularities can be handled by resumming $SU(N)$ propagators in
loops. Since the infrared structure of the theory is similar to commutative $SU(N)$ Yang-Mills,
thermal perturbation theory is expected to break down at order $\CO[(g^2 N)^3]$. 
As in the previous Section, for small $T\sqrt{\theta}$ we can obtain an asymptotic expansion in
powers of $T\sqrt{\theta}$. In the opposite limit we find that the $U(1)$ gauge boson
effectively interacts like a non-abelian Yang-Mills field in the ultraviolet.

This result extends to supersymmetric theories straightforwardly \refs\rgc\refs\rj. 
The Lagrangian of extended 
NCSYM theories can be written using the trick of dimensional reduction proposed in \refs\vm. 
We start with a $\CD_{\rm max}$-dimensional ``maximal'' $\CN=1$ theory on $\CM_{\theta}\times 
{\bf R}_{t} \times ({\bf S}^{1})^{\CD_{\rm max}-4}_{R}$ described by 
$$
\CL=-{1\over 2}{\rm Tr}\, F_{\mu\nu}\star F^{\mu\nu} + i {\rm Tr}\, 
\bar{\Psi}\Gamma^{A}\star D_{A}\Psi.
$$
As in \refs\vm\ we take the limit $R\rightarrow 0$ and retain only the zero modes of the 
fields in the internal coordinates, which in loop computations amounts to restrict momenta to 
four-dimensions in the Feynman integrals. Since all the internal coordinates are commutative,
the matrix $\theta_{ij}$ is completely oblivious of the fact that we are performing 
dimensional reduction. Thus, we get 
$\CN=1$, $\CN=2$ and $\CN=4$ NCSYM$_{4}$ by taking $\CD_{\rm max}=4,6,10$ respectively.

Again, Feynman rules for $U(N)$ NCSYM are retrieved by the replacement \rep\ on the rules for 
ordinary SYM. Repeating the analysis above, we find again that all dependence in $\theta$ 
comes from the contribution from $U(1)$ fields in non-planar diagrams, namely
$$
\eqalign{
\CF(T,\theta)_{\rm 2-loop}^{U(N)}
&= \CF(T,\theta=0)_{\rm 2-loop}^{SU(N)}+\half ({\CD}_{\rm max}-2)^{2}
\cr \times  & 
g^2N \int{d^{3}p\over (2\pi)^{3}}\int{d^{3}q\over 
(2\pi)^{3}} {1\over \omega_{q}\omega_{p}}\left[n_{b}(p)n_{b}(q)
+2n_{b}(p)n_{f}(q)+n_{f}(p)n_{f}(q)\right]
\sin^{2}\theta(p,q)
}
$$
where $\CF(T,\theta=0)_{\rm 2-loop}^{SU(N)}$ is the two-loop free energy density of $SU(N)$ 
ordinary SYM \refs\ft\refs\vm\refs\sjr\refs\an.

\newsec{Conclusions}

In the present paper we have studied several aspects of the thermodynamics of perturbative
gauge theories on noncommutative spaces. The case of $U(1)$ NCYM is the most interesting example from
a dynamical point of view. We have computed the quantum corrections to the one loop result and
found that, due to the restoration of the free theory at low momenta, the theory is free of infrared
divergences at three loops, due to the fact that the one loop two-point function of the gauge
field in the static limit vanishes quadratically with the external momenta. Although we did not 
explicitly compute corrections beyond order $\CO(g^2)$, the general structure of the theory 
seems to indicate that there is no onset of infrared divergences at any order. 

For large values of $T\sqrt{\theta}$ we have seen that the theory behaves in the ultraviolet as
an ordinary non-abelian Yang-Mills theory with $C_{2}(G)=2$. This is in perfect accordance with
the result of references \refs\rj\refs\rkw\refs\rcar, in the sense that the ultraviolet divergences of the
theory are obtained by averaging the factors $\sin^{2}\theta(p,q)$ in the amplitudes.
On the other hand, the theory is completely different from an ordinary non-abelian gauge theory
in the infrared. This is due to the fact that at large distances noncommutativity effects are
negligible and a free theory is restored. 

In the case of $U(N)$ NCYM, we have found that the two-loop free energy density only depends
on $\theta$ through terms of order $\CO(1)$ in the large-$N$ limit
that correspond to the contribution of 
the $U(1)$ part of $U(N)$ in non-planar loop diagrams. Since these $U(1)$ contributions are
regular in the infrared, we find that the infrared structure of the theory is determined by
the $SU(N)$ part and thus it is identical to that of ordinary YM theories. In particular, we will
have infrared singularities at three loops that can be resummed to give contributions of order
$\CO[(g^2N)^{3/2}]$ and $\CO[(g^{2}N)^{2}\log g^{2}N]$, as well as 
$\CO[(g^{2}N)^{5/2}]$. Thermal perturbation theory will break down at four loops
due to the self-interaction of magnetic $SU(N)$ gluons.

Here we have restricted ourselves to gauge theories on the noncommutative plane.
It would be interesting to better understand the case of $U(1)$ Yang-Mills theories in the noncommutative
torus, specially in the case of rational $\theta$. The computation of quantum corrections in this case is 
straightforward, and essentially amounts to replacing in our expressions momentum integrals by discrete
sums and rescaling $\theta_{ij}\rightarrow \pi\theta_{ij}$. By using Morita equivalence, 
one should be able to relate the free energy of these theories with that of ordinary gauge 
theories in the presence of twisted background fields.

\newsec{Acknowledgements}

We are pleased to thank Jos\'e Barb\'on, Jan de Boer, C\'esar G\'omez and Niels 
Obers for enlightening discussions. We also heartily thank Shiraz Minwalla for pointing out 
to us a mistake in the first version of this paper concerning Section 3. We are indebted to 
The Lorentz Center (Leiden University) and the organizers of the {\it Workshop on Noncommutative 
Gauge Theories} 
where part of this work was done. G.A. would like also to thank Spinoza Institute and 
especially Gerard 't Hooft for kind hospitality. The work of M.A.V.-M. has been supported by 
the FOM ({\it Fundamenteel Ordenzoek van the Materie}) Foundation 
and by University of the Basque Country Grants UPV 063.310-EB187/98 
and UPV 172.310-G02/99, and Spanish Science Ministry Grant AEN99-0315.

\listrefs

%\vfill\eject
\bye